\documentclass[conference]{IEEEtran}
\IEEEoverridecommandlockouts
\pdfoutput=1 
\usepackage[utf8]{inputenc}
\usepackage[T1]{fontenc}
\usepackage{amsmath,amssymb,amsfonts}
\usepackage{algorithmic}
\usepackage{graphicx}
\usepackage{textcomp}
\usepackage{xcolor}
\usepackage{multirow}
\usepackage{tabularx,ragged2e,booktabs,caption}
\usepackage{biblatex} %Imports biblatex package
\usepackage{enumitem}
\usepackage{pifont}
\def\BibTeX{{\rm B\kern-.05em{\sc i\kern-.025em b}\kern-.08em
    T\kern-.1667em\lower.7ex\hbox{E}\kern-.125emX}}
\usepackage{xspace}
\newcommand{\net}{QF-RobustNN\xspace}
\DeclareUnicodeCharacter{0301}{}
\usepackage{flushend}
\begin{document}

% package added by me
\newcommand{\todo}[1]{\textcolor{red}{#1}}
\newcommand{\wwnote}[1]{\textcolor{blue}{#1}}

\title{Can Noise on Qubits Be Learned in Quantum Neural Network? A Case Study on QuantumFlow\\
\vspace{5pt}

\vspace{-12pt}}
% Conference Paper Title*\\
\author{\IEEEauthorblockN{
Zhiding Liang\textsuperscript{\dag}, 
Zhepeng Wang\textsuperscript{\ddag},
Junhuan Yang\textsuperscript{\P}, 
Lei Yang\textsuperscript{\P},
Jinjun Xiong\textsuperscript{\S},
Yiyu Shi\textsuperscript{\dag},
Weiwen Jiang\textsuperscript{\ddag}}
\IEEEauthorblockA{
\textsuperscript{\dag}University of Notre Dame, IN, USA.
\textsuperscript{\ddag}George Mason University, VA, USA.\\
\textsuperscript{\P}University of New Mexico, NM, USA.
\textsuperscript{\S}University at Buffalo, NY, USA.\\
zliang5@nd.edu; wjiang8@gmu.edu
\vspace{-0.15in}}
}

\maketitle
\begin{abstract}
In the noisy intermediate-scale quantum (NISQ) era, one of the key questions is how to deal with the high noise level existing in physical quantum bits (qubits).
Quantum error correction is promising but requires an extensive number (e.g., over 1,000) of physical qubits to create one ``perfect'' qubit, exceeding the capacity of the existing quantum computers.
This paper aims to tackle the noise issue from another angle: instead of creating perfect qubits for general quantum algorithms, we investigate the potential to mitigate the noise issue for dedicate algorithms.
Specifically, this paper targets quantum neural network (QNN), and proposes to learn the errors in the training phase, so that the identified QNN model can be resilient to noise.
As a result, the implementation of QNN needs no or a small number of additional physical qubits, which is more realistic for the near-term quantum computers.
To achieve this goal, an application-specific compiler is essential: on the one hand, the error cannot be learned if the mapping from logical qubits to physical qubits exists randomness; on the other hand, the compiler needs to be efficient so that the lengthy training procedure can be completed in a reasonable time.
In this paper, we utilize the recent QNN framework, QuantumFlow, as a case study.
Experimental results show that the proposed approach can optimize QNN models for different errors in qubits, achieving up to 28\% accuracy improvement compared with the model obtained by the error-agnostic training.
\end{abstract}


\section{Introduction}
Quantum computing has the quantum potential to provide exponential speedup over the classical computers \cite{grover1996fast,tannu2018case,arute2019quantum,saha2020distributed,bertels2021quantum}, which is 
% As a result, quantum computing has become a hot area in recent years, and researchers 
considered to be one of the best candidates for solving some of the complicated computational problems that cannot be solved by classical computing in a reasonable time \cite{grover1996fast,xu2021variational,harrow2009quantum}. 
% Quantum computing can provide exponential acceleration and a very large computational capacity compared to classical computers. As a result, quantum computing has become a popular field in recent years and researchers consider it as one of the best candidates to solve some tricky and complex computational problems. 
We are now witnessing the rapid development in both quantum hardware and quantum algorithms: on the hardware side, the newly emerging quantum computers (such as IBM Q \cite{ibm2019experience}) are developed with superconductor and can perform superposition and quantum entanglement on the physical quantum bits (qubits); on the other side, different quantum algorithms have been devised, such as Shor's algorithm for primary factorization \cite{shor1994algorithms} and the quantum approximate optimization algorithm (QAOA)  for optimization problems \cite{farhi2014quantum}.
However, the current quantum computers have a high noise level and intermediate scale, known as noisy intermediate-scale quantum (NISQ).
How to deal with the noise with limited number of qubits becomes an imminent problem.

% The development of quantum computers, such as the IBM Q\cite{b9}, can simulate quantum computing and quantum phenomena such as superposition, quantum entanglement, and so on.

% Quantum algorithms have also been developed rapidly, and some excellent quantum algorithms such as the variational
% quantum eigensolver (VQE)\cite{b7}, the quantum approximate
% optimization algorithm (QAOA)\cite{b8} has been proposed. 

% The development of quantum computers, such as the IBM Q\cite{b9}, can simulate quantum computing and quantum phenomena such as superposition, quantum entanglement, and so on. Such modern quantum computers are defined as noisy intermediate-scale quantum (NISQ) hardware and are not fully fault-tolerant.

Quantum Error Correction (QEC) \cite{chiaverini2004realization} is a promising solution to solve the noise issue in quantum computing by creating ``perfect qubits'' (e.g., error rate to be less than $10^{-15}$).
It applies a set of noisy physical qubits to represent one perfect qubit, which can correct the errors made by the noise and significantly reduce the error rate.
Once the QEC is achieved, it can be beneficial for general quantum algorithms.
But, there is still a long way to achieve QEC, because it would require a large number (e.g., larger than 1,000 \cite{murali2019noise}) of physical qubits for a perfect qubit.
On the other hand, quantum computer in the NISQ era has a very limited number of qubits; for example, IBM Q has at most 65 qubits on the IBM-Manhattan backend. 
Therefore, it is critical to find a more efficient way to solve the noise issue in the NISQ era.

Instead of creating perfect qubits for general quantum algorithms with QEC, in this paper, we aim to address the noise issue in an application-specific way, such that no more qubits or a very small number of additional qubits are needed, while the effects of error can be mitigated. 
More specifically, we would like to explore the design of a dedicated quantum algorithm to make it resilient to noise.
Quantum neural network (QNN) is a good candidate to achieve such a goal since the neural network in computing-in-memory (CiM) platforms has already demonstrated its capability to be resilient to noise \cite{jiang2020device}.
Compared with the design of CiM-based neural network, quantum computing brings more challenges to identify the robust QNNs: (1) in terms of different properties (e.g., topology, error rate, number of qubits) on quantum computers, the implementations (or compilation) of the same QNN on hardware can be different,
% since the physical qubits are not fully connected, it requires designers to map QNN onto physical qubits and the mapping on different quantum computer can be different, 
making the learned model useless; (2) the mapping of QNN to quantum computer needs to be integrated into the training process, leading to
% in the learning phase, the mapping needing to be considered, which is time consuming and may easily become the performance bottleneck in the training algorithm, leading to 
lengthy training time; and (3) the update of one weight in the noisy quantum computer may affect other weights, resulting in the traditional training approach to be not applicable. The network structure can affects robust \cite{jiang2020standing,yang2020co,yang2020con,jiang2019achieving,jiang2019accuracy,song2021dancing, wang2021exploration, cheng2020accqoc,wang2021quantumnas}, which we will explore in the future.

In this paper, we base on a recent QNN design, i.e., QuantumFlow \cite{jiang2021co, jiang2021machine}, to demonstrate how to overcome the above challenges. 
We first devise an application-specific mapping algorithm.
The proposed approach can fix the mapping from logical circuits to physical qubits for different quantum computers; meanwhile, it can minimize the number of gates to be used in physical qubits.
In addition, the application-specific mapping follows the same procedure, which can significantly reduce the optimization time to better support training the neural network.
Then, based on the proposed mapping algorithm, we build up an error-aware training framework to identify the robust QNN model.
Kindly note that the training procedure is completed at classical computing in our implementation, which can be integrated into quantum computing when the scale of the quantum computer is ready for QNNs.

The contribution of this paper is three-fold as follows.
\begin{itemize}
    \item To our best knowledge, this is the very first work to learn the quantum error in quantum neural networks, and we demonstrate that a robust quantum neural network, namely \net, can be obtained.
    \item To support learning qubits' noise in quantum neural networks, we first devise an application-specific mapping method for one state-of-the-art quantum neural network, QuantumFlow.
    \item Then, a general training framework is proposed to conduct the error-aware learning, which can be easily extended to support training on the quantum computer in the future when the quantum computer is ready for quantum neural networks, in terms of the number of qubits.
\end{itemize}

The paper is organized as follows: In Section II, we present the preliminaries and motivations of this work. Section III presents our proposed mapping algorithm and training framework.
Evaluation results are reported in Section IV. Section V concludes this paper.

\begin{figure}[t]
\centerline{\includegraphics[width=1\linewidth]{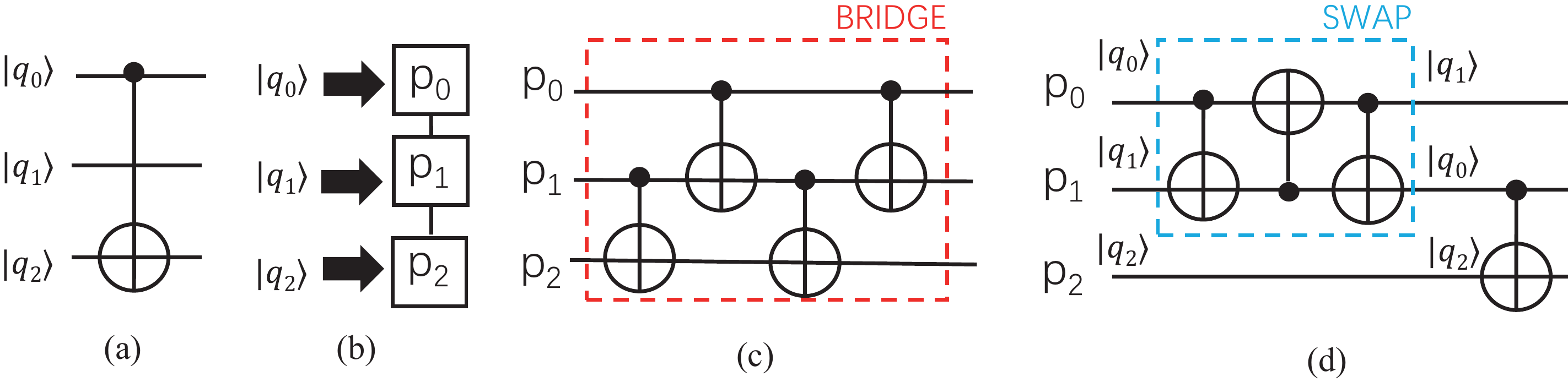}}
\caption{BRIDGE and SWAP: (a) A CNOT gate with distance 2. (b) Assumed physical qubits connection. (c) Use the BRIDGE gate to achieve gate (a). (d) Use SWAP gate to achieve gate (a).}
\label{fig:Swapbridge}
\end{figure}

\section{Preliminaries}
% \subsection{Quantum Basics}
\textbf{Qubits}: In classical computing, the basic unit to represent data is a binary bit (i.e., 0 or 1); similarly quantum computing utilizes quantum bit (qubit) for data representation.
Unlike a classical bit that can only represent 0 or 1, the qubit is to represent 0 and 1 simultaneously.
More specifically, it describes a state $|\psi\rangle$ with the combination of the 0-state (i.e., $|0\rangle$) and 1-state (i.e., $|1\rangle$); i.e., $|\psi\rangle=\alpha|0\rangle+\beta|1\rangle$,  where $\alpha$ and $\beta$ are the amplitudes of 0-state and 1-state, and $\alpha^2$ and $\beta^2$ represent their probabilities.
% We have $\alpha^2+\beta^2=1$.
When $\alpha\ne 0$ and $\beta\ne 0$, the qubit can represent two-state at the same time, known as the ``superposition'' state.

% In classical computers, there is the basic unit of the bit, which is expressed in quantum computing using a similar quantum bit (qubit). Similar to how a classical bit can be in a 0-state or a 1-state, a qubit has a similar state. However, the state of a bit is a number (0 or 1), while the state of a qubit is a vector. More specifically, the state of a qubit is a vector in a two-dimensional vector space. This vector space is called the state space. A qubit can be in states $|0\rangle$ and $|1\rangle$ and also in linear superposition states of $|0\rangle$ and $|1\rangle$, which in physics we call quantum superposition states. $|\psi\rangle$ = $\alpha|0\rangle$ +$\beta |1\rangle$.Here $\alpha$ and $\beta$ are both complex numbers and 
% $|\alpha|^2+|\beta|^2=1$.
% This two-dimensional complex vector space gives it the ability to 'superposition', and this 'superposition' has an infinite number of possibilities.\\
\textbf{Quantum Gates}: In analogy to the classical logic gates that process classical data, quantum gates manipulate quantum information, i.e., the quantum states qubits.
A quantum system is typically initialized to zero state (i.e., $|\psi\rangle=|0\rangle$).
Given a Hamiltonian, \cite{griffiths2018introduction} has demonstrated that the Hamiltonian can be solved by Schrödinger's equation, which can be decomposed into various combinations of single-qubit (e.g., H, S, and T gates) and multi-qubit quantum gates (e.g., CNOT gate).
It has been proven that the combination of Clifford gates (e.g.,  CNOT, H, S gates) and T gate is a universal gate set to represent arbitrarily complex gates for quantum computing.

% Theoretical calculations show that the set of single-qubit gates (e.g., ) and multi-qubit gates CNOT can constitute arbitrarily complex gates for quantum computing.

% A quantum computing system, if its initial state $|\psi_t = 0\rangle$, and Hamiltonian are known, can be known by solving Schrödinger's equation for any moment state $|\psi_t\rangle$. Since the state of a single quantum bit will be initialized at $|0\rangle$, i.e., the initialized state of the system is generally deterministic, the most important thing is to construct the Hamiltonian of the system. The evolution of the Hamiltonian can be decomposed into various combinations of single-qubit and multi-qubit quantum gates. 

% In analogy to the classical logic gates used in ordinary everyday computers, such as with, or and without gates, the complex quantum computation can also be achieved by decomposing into a series of similar combinations of quantum gates. Theoretical calculations show that the set of single-qubit gates and multi-qubit gates CNOT can constitute arbitrarily complex gates for quantum computing.

\textbf{Quantum Circuit}: A series of logical gates can construct the classical circuit; likewise, a series of quantum gates form the quantum circuit. 
According to the function/algorithm needing to be implemented, a quantum circuit can be is formed by using the quantum gates in the universal gate set.
The length/depth of the circuit reflects the time complexity of the quantum version of the implemented algorithm, and the width of the circuit indicates the number of qubits to be used.

% Also known as quantum logic circuits, are the most commonly used generic quantum computing model. It represents the line in which operations are performed on qubits under the abstract concept. It is formed by a series of single or multiple qubit gate operations being applied to a qubit to perform some quantum algorithm, and the result is often finally read out by the need for quantum measurements. In a quantum circuit, the circuit is connected by time, and the process is operated as directed by the Hamiltonian operator until it meets a quantum gate. Since each quantum gate is a unitary operator, the whole quantum circuit as a whole is a large unitary operator.

\textbf{Quantum Processor}: The quantum processor (a.k.a., quantum computer) is built upon a set of physical qubits, which are implemented by superconductor \cite{monroe2014large}, IoN-Trap \cite{monroe2013scaling}, etc.
Taking the commonly superconductor-based quantum processor as an example, it has two limitations: (1) qubits are not fully connected; and (2) noise exists in operating qubits or a pair of qubits.
% Different techniques are proposed to overcome these limitations.

\begin{table}[t]
  \centering
  \tabcolsep 6pt
  \caption{Accuracy in Different Noise Models}
    \begin{tabular}{|c|c|c|c|c|}
    \hline
    Noise Model &
      Affected Gates &
      Error Rate &
      Accuracy &
      Time
      \\ \hline
    No Error &
      0 &
      - &
      0.9804 &
      5.00s
      \\\hline
    Flip Error &
      X, CX, CCX &
      0.1 &
      0.5333 &
      568.50s
      \\\hline
    Flip Error &
      X, CX, CCX &
      0.01 &
      0.8824 &
      540.00s
      \\\hline
    Phase Error &
      Z, CZ &
      0.1 &
      0.6429 &
      545.00s
      \\\hline
    Phase Error &
      Z, CZ &
      0.01 &
      0.9167 &
      511.14s
      \\\hline
    Flip+Phase &
      X,CX,CCX,Z,CZ &
      0.1 &
      0.451 &
      628.00s
      \\\hline
    Flip+Phase &
      X,CX,CCX,Z,CZ &
      0.01 &
      0.7778 &
      532.04s
      \\\hline
    \end{tabular}%
  \label{tab:noise}%
\end{table}%

\textit{SWAP and BRIDGE for Limitation (1)}. Figure \ref{fig:Swapbridge} shows an example with 3 qubits organized in a linear form, where logical qubit $|q_i\rangle$ is mapped to physical qubit $p_i$, i.e., $M(|q_i\rangle)=p_i$.
And a CNOT gate operates between $M(|q_0\rangle)$ and $M(|q_2\rangle)$.
Obviously, we cannot directly conduct CNOT the physical qubits has no channel for communication.
To solve this, a BRIDGE operation and a SWAP operation in Figure \ref{fig:Swapbridge}(c)-(d) are designed.
The idea of BRIDGE in Figure \ref{fig:Swapbridge}(c) is to create a unitary operator that performs the function of CNOT gate between two physically separate qubits, while the state of the middle qubit (i.e., $M(|q_1\rangle)$) is not changed.
On the other hand, the SWAP in Figure \ref{fig:Swapbridge}(d) is to exchange the quantum states between $p_0$ and $p_1$, as such, after the SWAP, we will have $M(|q_0\rangle)=p_1$ and $M(|q_1\rangle)=p_0$, which can perform CNOT on $|q_0\rangle$ and $|q_1\rangle$.
It is worth noticing that both BRIDGE and SWAP can work for the communication between two qubits needing to cross more than 1 physical qubit, but needing more CNOT gates for implementation, introducing higher cost.

\textit{Noisy Qubits leading to Limitation (2)}: Due to the physical material, there exists a high noise level in quantum qubits. 
In classical computing, the ``bit flip'' is a typical error; when it comes to the qubit, there are more types of error, which can be concluded as follows:
% \textbf{Quantum Error} - This is the noise interference that would be encountered when a quantum circuit is run on a real quantum computer. Quantum errors can be classified as the list in the following cases depending on different kinds of impact:

\noindent \textit{(1) Qubit-flip Error}: This is manifested as a state flip of a qubit from $|0\rangle$ to $|1\rangle$ or versus. 

\noindent \textit{(2) Phase Error}: This is manifested as a phase shift of a qubit, which can be regarded as the flip on the phase (e.g., perform a Z gate). 

\noindent \textit{(3) Readout Error}: This is specified by a list assignment probability vector $P(n|m)$, assuming $m$ is a bit string from an ideal measurement, while $n$ is from a noisy measurement. $P(n|m)$ reflects the number of errors exists in $n$.

\noindent \textit{(4) Depolarizing Error}: An n-qubit channel stores depolarization error parameterized by a depolarization probability $p$.

% which reflects the probability that the number of bits 

% n is a noisy measurement given a true ideal measurement of m, where n and m are integer representations of the bit string.

% % from $|1\rangle$ to $|0\rangle$. The function is similar to an extra X gate acting on the quantum circuit.

% \begin{itemize}[noitemsep,topsep=0pt,parsep=0pt,partopsep=0pt]
% \item Qubit-flip Error - A Pauli error which is manifested as a state flip of a qubit, e.g., from $|0\rangle$ to $|1\rangle$ or from $|1\rangle$ to $|0\rangle$. The function is similar to an extra X gate acting on the quantum circuit.
% \item Phase Error - A Pauli error is manifested as a phase shift of a qubit. The function is similar to an extra Z gate acting on the quantum circuit.
% \item Readout Error - Is specified by a list assignment probability vector $P(A|B)$. This is the probability that n is a noisy measurement given a true ideal measurement of m, where n and m are integer representations of the bit string.
% \item Depolarizing Error - An n-qubit channel stores depolarization error parameterized by a depolarization probability $p$.
% \end{itemize}

% which is considered as noise models with great impact on the operation of quantum circuits.

\textbf{Quantum Compiler}: In order to implement a quantum circuit to a specific quantum process, a compiler is essential to map the logical circuit to physical qubits with the different objectives (e.g., minimizing the number of quantum gates, circuit depth, fidelity, etc.).
The existing quantum compiler is almost designed for general purpose \cite{cowtan2019qubit}, but we observe that they may be too costly to be applied for the scenario needing to compile a large number of logical circuits.
And this work aims to provide an application-specific quantum compiler to find the most efficient mapping.

\textbf{Target Quantum Application}: The application considered in this work is Quantum Neural Network (QNN), which is recently emerged and attacked wide attention from both academics and industry.
In the training phase of QNN, it may require compiling an extensive number of quantum circuits, which has a high demand on the compiler efficiency.
For example, FFNN \cite{tacchino2020quantum} and QuantumFlow \cite{jiang2021co} will build different quantum circuits for different weights.

% In this work, we focuses on phase error, bit-flip error, and their combined phase-bit flip error, but other errors can be easily integrated in the proposed model.

% Is a co-design framework for neural networks and quantum circuits, following the network/hardware co-design philosophy\cite{jiang2019accuracy}\cite{jiang2020hardware}, which considers the implementation of quantum circuits from the circuit perspective in the design process of neural networks and the network properties in the synthesis of quantum circuits. QuantumFlow contains five components (QF-pNet, QF-that, QF-FB, QF- Circ, and QF-Map), which work in concert to design neural networks that represent data as random variables to exploit quantum capabilities\cite{jiang2021co}.

\begin{figure}[t]
\centerline{\includegraphics[width=1\linewidth]{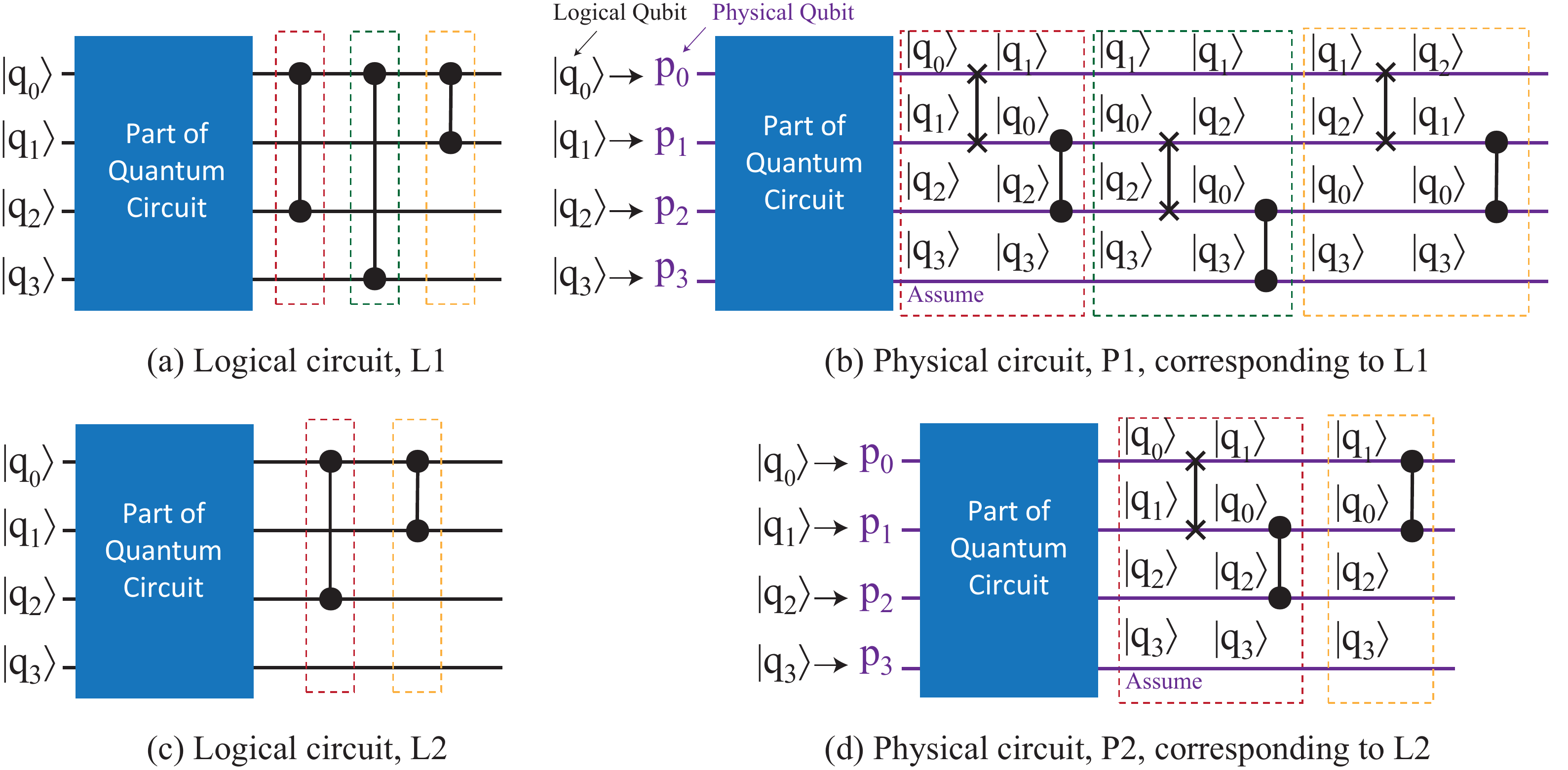}}
\caption{Illustration of error unpredictable caused by compilation: (a) and (c) shows two logical quantum circuits L1 and L2. (b) and (d) describe the steps processing in the physical circuit P1 and P2, which corresponding to L1 and L2, respectively.}
\label{fig:gate}
\end{figure}

\section{Motivation}
% \subsection{Quantum Error-aware Problem}
% The implementation of quantum computing faces a fundamental difficulty\cite{wille2019mapping}. As with other processes in nature, noise interference is everywhere. When a quantum circuit is placed in a different noise model, even small phase errors and flip errors still have a significant impact on the accuracy of the results. 
Noise can significantly affect the performance of QNN.
In Table \ref{tab:noise}, we report the results on the MINST database: QuantumFlow on perfect qubits can obtain a 98.04\% of accuracy. When we add bit-flip errors with error rates of 0.1 and 0.01, the accuracy degraded to 53.33\% and 88.24\%, respectively; when we add phase errors with error rates of 0.1 and 0.01, the accuracy is 64.29\% and 91.67\%, respectively
Moreover, for a quantum processor with both bit-flip and phase-error and the error rate to be 0.1 and 0.01, the accuracy further reduced to 45.10\% and 77.78\%, respectively.
From the above results, it is easy to see how noise interferes with QuantumFlow.
With such observation, we are thinking of whether it is possible to learn the error in the training phase, such that we can obtain a more robust QNN.

% However, with the addition of an error-aware compiler, the interference of these noises will be greatly reduced.

% are added to the 'X', 'CNOT', and 'CCNOT' gates of all qubits, the accuracy of the QuantumFlow optimized circuit reduces to 53.33\% and 88.24\%. When the phase errors with error rates of 0.1 and 0.01 are added to the 'Z', and 'CZ' gates of all qubits, the results of the QuantumFlow optimized circuit become 64.29\% and 91.67\%, respectively. Then when the phase error and bit-flip error are composed together and added to the 'X', 'Z', 'CNOT', 'CZ', 'CCNOT' gates with error rates of 0.1 and 0.01, respectively, the accuracy of the QuantumFlow optimized circuit becomes 45.10\% and 77.78\%. It is easy to see how noise interferes with the QuantumFlow optimization circuit. However, with the addition of an error-aware compiler, the interference of these noises will be greatly reduced.

We have seen the success of learning errors in classical computing \cite{havlivcek2019supervised}.
But, when it comes to quantum computing, more challenges pose up.

\textbf{Challenge 1: Error unpredictable on the quantum circuit.}

In quantum processors, the error on qubits is usually different; if a weight in QNN cannot be mapped to a fixed qubit for execution at run-time, it will lead to the error being unpredictable.
Consider a QNN with two different weights, which are embedded into the circuit with different quantum circuits (L1 in Figure \ref{fig:gate}(a) and L2 in \ref{fig:gate}(c)).
% \subsection{Challenges and Observations}

% \subsubsection{Error unpredictable on quantum circuit challenge}
% Determined by the nature of some kinds of quantum frameworks, the quantum circuit generated by each different training weight is different. 
% And the error information contained in the quantum circuit may affected by the prior quantum circuit that generated by the prior training weight. 
% For example, a logical quantum circuit L1 shows in the Figure \ref{fig:gate}(a). And in 
For L1, Figure \ref{fig:gate}(b) shows its implementation on the quantum process in Figure \ref{fig:Swapbridge}(b).
% gates in different color's dotted box corresponding to gates in same color dotted box in Figure \ref{fig:gate}(a). 
Assume the logical qubits ($q_0, q_1, q_2, q_3$) are mapped to the physical qubits as ($p_0, p_1, p_3, p_4$) after the data pre-processing (blue box). 
In embedding the weights into the circuit (last CZ gate with two dots), qubit $q_0$ is on physical qubit $p_2$, while $q_1$ is on $p_1$.
On the other hand, due to the change of weights, L2 does not have the second CZ gate (blue dashed box).
As a result, qubit $q_0$ is on physical qubit $p_1$, while $q_1$ is on $p_0$.
As a result, when we execute the circuit, the error for the same weight can be totally different in two QNN models.
This will hinder us to learn the error in the QNN model.

\begin{table}[t]
  \centering
  \tabcolsep 9pt
  \caption{Compilation time of QUEKO \cite{tan2020optimal}.}
    \begin{tabular}{|c|c|}
    \hline
    Quantum Gate(with 4 qubits) &
      Compilation Time
      \\ \hline
    CCX &
      1.73s
      \\ \hline
    CZ &
      0.14s
      \\ \hline
    CCX + CZ +CCX &
      16.43s
      \\ \hline
    CCX + CZ +CCX + CCX &
      91.91s
      \\  \hline
    CCX + CZ +CCX + CCX + CZ &
      128.20s
      \\ \hline
    CCX + CZ +CCX + CCX + CZ + CCX &
      439.67s
      \\ \hline
    \end{tabular}%
  \label{tab:runtime}%
\end{table}%

\textbf{Challenge 2: Existing compiler may not for training.}

% \subsubsection{Qubits mapping runtime problem}
% The traditional mapping depends on the state of the previous gate. In some sorts of quantum frameworks, the previous gate , so the information behind it may change, and it is not sure whether it is the same as the initial state, e.g. QuantumFlow.
It is well known that compiling is an NP-Hard problem.
For a general-purpose compiler, it will have a high execution time for better exploring the search space.
On the other hand, since we need to integrate the compiling into the training iteration, and it is also well known that training in the neural network is extremely time-consuming.
% In traditional mapping methods, a new quantum circuit must be mapped once. It's one of the reason that the compiling of quantum circuits takes a long time.
% The problem of error unpredictable in the paper also needs to consider the error information training into the new quantum circuit model. 
% However, a large runtime is not doable for Noisy Intermediate-Scale Quantum(NISQ). 
In terms of compilation time, we exam one of the state-of-the-art compilers \cite{tan2020optimal} with the configuration of ``fidelity'' and ``transition'' (a high-speed processing mode).
% , which constructs the benchmark in reverse according to the structure of the quantum device. 
Table \ref{tab:runtime} reports the result: the compilation time increases significantly with the increase of quantum gate numbers in the quantum circuit.
More specifically, it takes less than 1 second for one gate, but the compilation time grows to more than 400 seconds for 6 gates.
This is commonly existing different general-purpose compiler since the increase of the number of gates can exponentially increase the search space, leading the compilation time increase exponentially.
However, the quantum circuit in QuantumFlow is far more complicated than the circuit with 6 gates in Table \ref{tab:runtime}, and we need to conduct such compilation each time for a new weight is identified in the training phase.

% First construct a quantum circuit with known optimal solutions, and then generate a random mapping of physical qubits to logical qubits to obtain a quantum circuit that does not satisfy the physical constraints, which acts on many details of qubits mapping. However, in complex quantum systems, this approach has a disadvantage in terms of running speed because compiler deals with depth details. This property makes this compiler unsuitable for complex quantum circuits such as the quantum circuits generated by Quantumflow. After trying to test the performance of this compiler on some given quantum circuits using "fidelity" and "transition" (a high-speed processing mode in this compiler), the compilation time increases significantly with the number of quantum gates in the quantum circuit, as shown in the Table \ref{tab:runtime}. And after that the compilation time changes exponentially with the number of quantum gates. This time consumption clearly shows that this compiler is not suitable for very complex quantum circuits, while ours can be scaled down to meet the requirements.

\textbf{Motivation}: After summarizing the above-described observations and challenges, it is clear that we need to resolve the effects of qubits' error to make QNN effective, and we need to reduce the compilation time to make it possible to learn the error in the training phase.

% There is sufficient motivation for this mission: Give a application-specific qubits mapping algorithm to fix the mapping from logic qubits to physical qubits, so that error can be predicted. The method should also make the actual cost of the quantum circuit less and also to improve the efficiency.

\section{Proposed Algorithm}
\subsection{Problem Statement}
With the demonstration of the existing challenges and our observations, it is clear that the existing errors in quantum computers will greatly reduce accuracy.
In addition, in order to learn the error in the QNN mode, the compilation should be more efficient to support the training.
% The training itself is time-consuming and cannot use simulation to incorporate errors into the training procedure.
We formally define the problem of `learning quantum error in QNN'  as follow:

\begin{figure}[t]
\centerline{\includegraphics[width=0.7\linewidth]{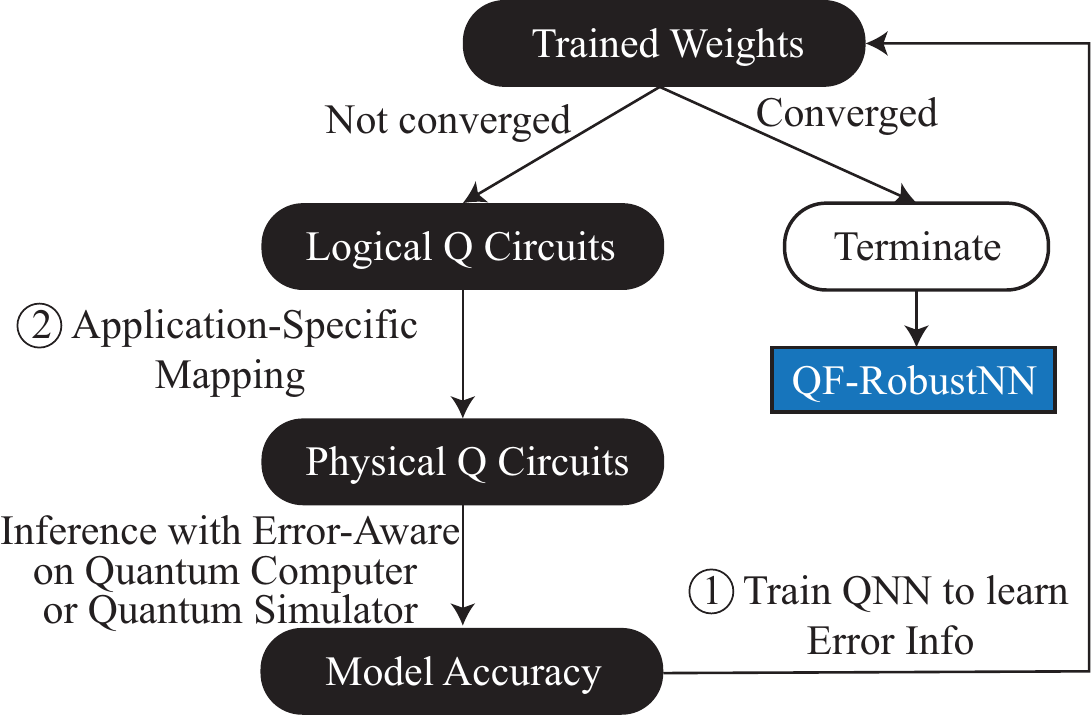}}
\caption{Illustration of the proposed training framework to generate \net.}
\label{fig:process}
\end{figure}

\textbf{Quantum-error-aware neural network training:} Given a neural network and its implementation on physical qubits, how to search for the weights of the neural network, such that the accuracy of the neural network model can be maximized with the consideration of error on qubits.

In the above problem, the implementation of a given quantum circuit needing to be implemented by a compiler, and therefore, we will further study how to design an application-specific compiler to support the training procedure, which is defined as follows.

\textbf{Application-specific compiler:} Given a dataset, the quantum algorithm, and quantum processor, how to map logical circuit to the quantum processor, such that each qubit is fixed to be mapped to a physical qubit and the circuit length can be minimized.\\

% The quantum gates generated according to different training weights are important for the whole experiment, and quantum neuron network are used to search the best performing training weights in that state and then update the training weight to improve the accuracy.\\
% \textbf{Objectives:} As stated, our objectives include: the qubit mapping from logic qubits to physical qubits needing to be fixed, since the qubit mapping algorithm is invoked by the training model in the QNN, the error can be learned in the QNN. A hybrid swap-bridge circuit method and prune branches to accelerate the simulation.

\subsection{Quantum-error-aware training framework}
In this section, we provide detailed procedures for the operation of the quantum-error-aware training framework, as shown in Figure \ref{fig:process}.
The framework is composed of four components: \textit{i)} the start point with trained weights; \textit{ii)} the logical quantum circuit created by QuantumFlow; \textit{iii)} the mapping from logical circuit to physic qubits; and \textit{iv)} complete the inference to obtain model accuracy and tune the parameters accordingly.
If the model is converged after the training step, we will terminate the algorithm and output the identified \net.

To better demonstrate the framework, we formulate the procedures as follows:
\begin{equation}
\begin{aligned}
c_i = circ(W_i)\\
% E(W_i)\\
m_i = Map(c_i, PhyQ)\\
e_i = Error(m_i)\\
a_i = Inference(m_i, e_i)
\label{eq1}
\end{aligned}
\end{equation}
% \textbf{Input:} The first step is to input the target quantum circuit as well as the quantum algorithm.\\
For one identified weight in the $i^{th}$iteration, the circuit $c_i$ is obtained, where the $circ$ represents obtain the circuit from a QNN framework (e.g., QuantumFlow).
Then, we map the circuit to the physical qubits (PhyQ) with a mapping function $Map$, which will be introduced in the next subsection.
After this, we will build up the error model $e_i$ based on the mapping $m_i$.
Finally, we will conduct the inference to obtain the accuracy ($a_i$) of the QNN with weight $W_i$, which is obtained by executing the physical circuit $m_i$ with the error model $e_i$.
After this, we will be going through the training procedure to update the weights in the QNN.

% conduct the inference phase 

% an error function $E()$

% In the second part, according to the \eqref{eq1}, an error function $E()$ containing error information is built according to the following steps: given a training weight parameter $W_i$, a quantum circuit $c_i$ is obtained according to this parameter $W_i$. $m_i$ is obtained by mapping logical qubits to physical qubits according to the qubits mapping method in our compiler. finally, the error function $E()$ is built by simulating $m_i$ with the parameter error.\\

\textbf{\ding{192} Train QNN to learn error info:} In the third step in Figure \ref{fig:process}, the QNN training model is built up with the purpose of searching for the best training weights is created.
The training weight will be learned with the error function $Error$ to become be aware of the quantum error and make the model resilient to the error $Error$. 
Given an initial weight that performs best in perfect quantum qubits, the training procedure is to update the weights.
Here, we can apply different optimization approach to update the weights. For example, we can apply deep reinforcement learning, where a recurrent neural network (RNN)-based controller will guide the updates.
More specifically, the accuracy $a_i$ will be the input of the controller to update the parameters of the RNN, and the RNN will then generate the weights to update the training weights of the first block in Figure \ref{fig:process}.
% by randomly generating different training weights back to the quantum circuit in the quantum mapping approach to obtain new results. 
The best performing weight is denoted as 'Searched\_Weight' by choosing the highest accuracy among all results. Comparing the results of 'Searched\_Weight' with the results of the initial weights (called baseline weights), we can evaluate the effectiveness of the proposed training framework, which will be reported in Section V.
% the improvement of the training model on accuracy can be known.

% \usepackage{graphicx}

% \usepackage{multirow}

\begin{figure}[t]
\centering
\centerline{\includegraphics[width=1\linewidth]{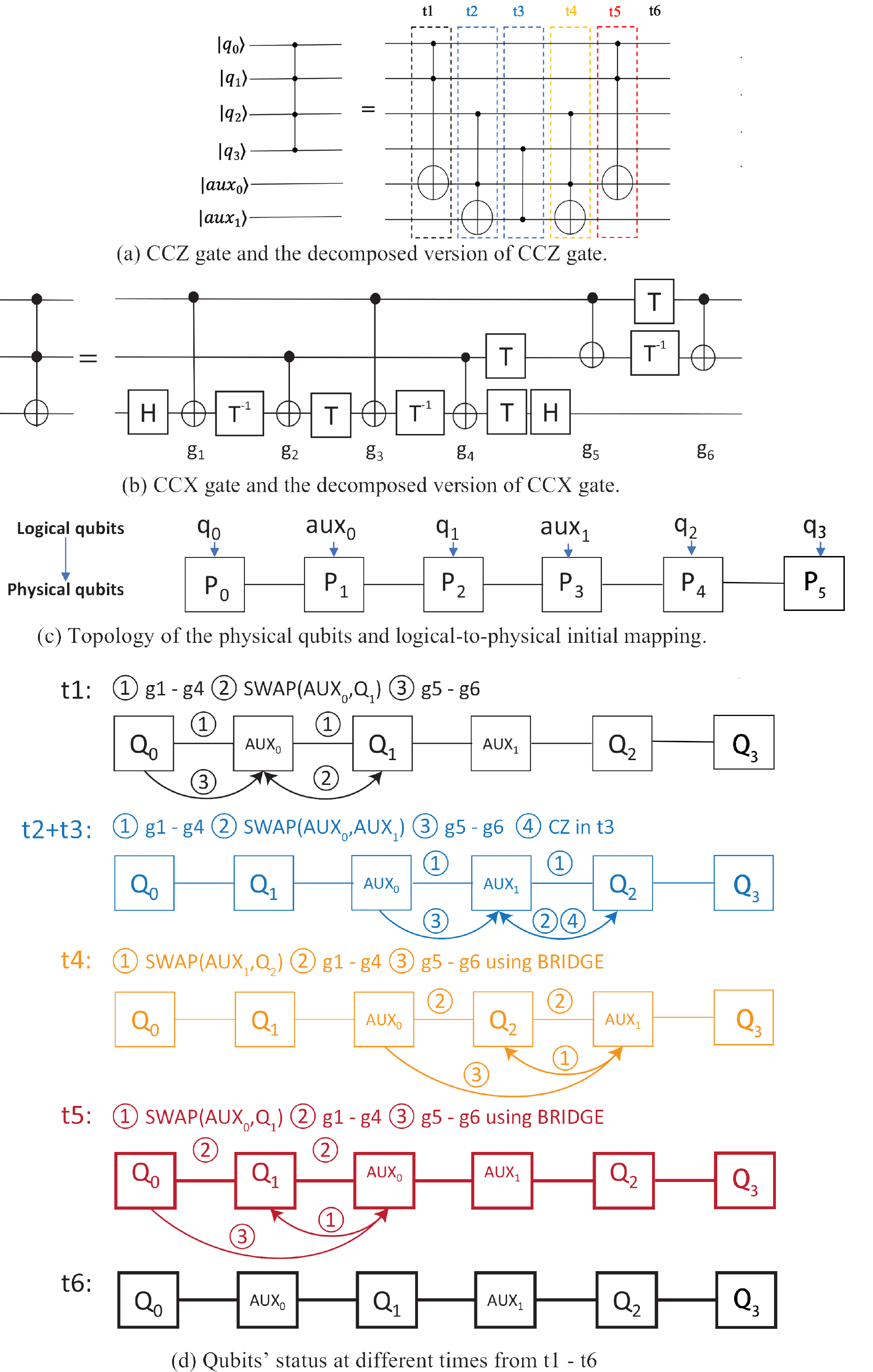}}
\caption{Describe the proposed qubits mapping algorithm steps and the corresponding gate connective behavior. (a) CCCZ gate and the decomposed version of CCCZ gate. (b) CCX gate and the decomposed version of CCX gate. (c) is the assumed physical qubit coupling graph and the step-by-step operations of the proposed qubit mapping algorithm.}
\label{fig:mapping}
\end{figure}

\textbf{\ding{193} Application-specific Mapping}
$C^{N}Z$ gates are the common gates utilized by QuantumFlow to implement the quantum neural networks, where $N$ represents there are $N$ control end with one target qubits using $Z$ gate.
To simplify the presentation, we choose $C^3Z$ (i.e., CCCZ) gates as an example to describe our application-specific qubit mapping algorithm.

The CCCZ gate and the decomposed circuit of the CCCZ gate are illustrated in Figure \ref{fig:mapping}(a). 
Then, the decomposed circuit of the CCX gate which shows in Figure \ref{fig:mapping}(b) that consists of basic CNOT gates to be executed during CCX gate operation that revealing logic qubits connection relationship. 
The topology of the physical qubits and logical-to-physical qubits initial mapping is provided in Figure \ref{fig:mapping}(c).
The assumed distribution and connectivity of physical qubits are shown in the coupling graph, where $Q_n$ refers to the logic qubits $q_n$ corresponding to their physical qubits in the quantum circuit and $AUX_n$ refers to the auxiliary qubits $aux_n$ corresponding to their physical qubits in a quantum circuit.

To make the qubit mapping for different $C^NZ$ gates to be fixed, our design philosophy is to make the logical-to-physical mapping to be the same at the beginning and end of the mapping procedure, meanwhile, we do not incur additional SWAP gates to minimize the circuit length. The procedure of the mapping algorithm is shown in Figure \ref{fig:mapping}(d). The dotted boxes with the same color in Figure \ref{fig:mapping}(a) and Figure \ref{fig:mapping}(d) have correspondence.  In the first step, we need first to conduct the qubit level logical-to-physical mapping. This will be followed by the rule of interleaving the computing qubits and auxiliary qubits, and they are mapped to a chain of qubits. For example, Figure \ref{fig:mapping}(c) gives an initial logical-to-physical mapping: $\{q_0, aux_0, q_1, aux_1, q_2, q_3\} \rightarrow \{P_0, P_1, P_2, P_3, P_4, P_5\}$. For convenient corresponding understanding, we refer $\{P_0, P_1, P_2, P_3, P_4, P_5\}$ as $\{Q_0, AUX_0, Q_1, AUX_1, Q_2, Q_3\}$ in Figure \ref{fig:mapping}(d). Then, the computation of CCX gates will be arranged in two-phase: forward phase and backward phase.

In forward phase, at $t_1$, the CCX gate in black dotted box act on $q_0$, $q_1$ and $aux_0$ in Figure \ref{fig:mapping}(a). The corresponding  decomposed circuit in Figure \ref{fig:mapping}(b) should be executed. It is obvious to see in the coupling graph in Figure \ref{fig:mapping}(c) that $Q_0$ has already been connected to $AUX_0$, and $AUX_0$ is already connected to $Q_1$ by an edge of the coupling graph in black dotted box, so that $g_1$, $g_2$, $g_3$, and $g_4$ in Figure \ref{fig:mapping}(b) can be directly executed. However, we still need to build connection between $Q_0$ and $Q_1$. Thus, we insert a SWAP operation SWAP$(Q_1, AUX_0)$ before $g_5$ to move logic qubit $q_1$ toward $aux_0$. After the SWAP operation, the mapping is update to $\{q_0, aux_0, q_1, aux_1, q_2, q_3\} \rightarrow \{Q_0, Q_1, AUX_0, AUX_1, Q_2, Q_3\}$, and $Q_0$ is connected to $Q_1$ that $g_5$ and $g_6$ can be executed. At $t_2$, a same procedure as $t_1$ is performed, and finally we can see that $|Q_1\rangle$ and $|AUX_1\rangle$ have connection in physical qubits, and we can perform the CZ gate in place at $t_3$. 

In backward phase, operations are processed at $t_4$ and $t_5$ in Figure \ref{fig:mapping}(c). At $t_4$, The required connection relationship is $Q_2$ connected to $AUX_1$, $AUX_0$ connected to $AUX_1$, and $Q_2$ connected to $AUX_0$. First step is put a SWAP operation SWAP$(Q_2, AUX_1)$ in yellow dotted box before executing $g_1$ to move logic qubit $q_2$ toward $aux_1$. And the mapping is update to $\{ q_0, aux_0, q_1, aux_1, q_2, q_3\} \rightarrow \{Q_0, Q_1, AUX_0, AUX_1, Q_2, Q_3\}$. This step is in order to build connection between $AUX_0$ and $AUX_1$, $Q_2$ and $AUX_1$ by the edge of the coupling graph in yellow dotted box in Figure \ref{fig:mapping}(c), so that $g_1$, $g_2$, $g_3$ and $g_4$ can be executed. The $AUX_0$ is needed to connect with $Q_2$. Here we do a BRIDGE operation BRIDGE$(Q_2, AUX_0)$ in order to save the amounts of basic CNOT gates, because if we do twice SWAP operation here, the cost number of basic CNOT gates will be 8. BRIDGE operation could be only using 3 basic gate here due to the middle logic qubit $|aux_1\rangle$ could be in state $|0\rangle$. And after BRIDGE operation, $AUX_0$ is connected with $Q_2$ via the edge of bridge, $g_5$ and $g_6$ in yellow dotted box can be executed. At $t_5$, the same procedure is repeated.
As a result, the qubit level logical-to-physical mapping is resumed, so that we can achieve our goal to make all weights to be mapped in a fixed and predictable way.

% there is a similar condition that a SWAP operation SWAP$(AUX_0, Q_1)$ is applied to update mapping to $\{ q_0, aux_0, q_1, aux_1, q_2\} \rightarrow \{Q_0, AUX_0, Q_1, AUX_1, Q_2\}$ and a BRIDGE operation BRIDGE$(Q_0, Q_1)$ is added to build connection between $Q_0$ and $Q_1$ in red dotted box in Figure \ref{fig:mapping}(c). The final mapping is $\{ q_0, aux_0, q_1, aux_1, q_2\} \rightarrow \{Q_0, AUX_0, Q_1, AUX_1, Q_2\}$ which is same as the initial mapping, the logical-to-physical qubit mapping is fixed and the error can be predictable.

\section{Results and Discussions}
This section describes the experimental setup and results. We are given an initial weight that performs best on a synthetic dataset in the perfect model as our baseline for accuracy comparison.
% The weights are then updated by searching for new training weights following the procedure in Figure \ref{fig:process}.
% After a given number of training iterations, the best weights with best performance will be identified as our ``search weights''. 
In addition, the existing compilers will be the baselines for efficiency comparison.
% carried out to demonstrate the efficiency of our proposed approach.

\begin{table}[t]
  \centering
  \tabcolsep 1.57475pt
  \caption{Accuracy Test Result From Proposed Algorithm in Different Models}
    \begin{tabular}[\linewidth]{|c|c|c|c|}
    \hline 
    \multirow{2}[0]{*}{Error Rate} &
      Baseline &
      \multicolumn{2}{c|}{\net}
      \\ \cline{2-4}
     &
      Acc. &
      Acc. &
      \multicolumn{1}{c|}{Weight}
      \\ \hline
    0 (Perfect) &
      94\% &
      - &
      [-1,-1,1,1,1,1,1,1],  [1,1,1,1,1,1,1,1]
      \\ \hline
    0.0001 &
      84\% &
      84\% &
      [-1,-1,1,1,1,1,1,1],  [1,1,1,1,1,1,1,1]
      \\ \hline
    0.0005 &
      75\% &
      76\% &
       [1,1,1,-1,-1,-1,-1,-1], [-1,-1,-1,1,1,-1,-1,-1]
      \\ \hline
    0.001 &
      74\% &
      76\% &
       [1,1,-1,1,1,-1,-1,1], [-1,1,1,1,-1,1,1,1]
      \\ \hline
    0.01 &
      75\% &
      80\% &
      [-1,-1,1,-1,1,-1,1,1],  [1,1,1,-1,-1,1,-1,1]
      \\ \hline
    0.05 &
      49\% &
      75\% &
       [1,1,-1,-1,1,-1,1,1], [-1,1,-1,-1,-1,-1,-1,1]
      \\ \hline
    0.1 &
      47\% &
      75\% &
       [1,1,-1,-1,1,-1,1,-1],  [1,1,-1,-1,1,-1,1,-1]
      \\ \hline
% \multicolumn{4}{l}{\footnotesize *Baseline\_Weight = [-1, -1,  1,  1,  1,  1,  1,  1],  [1,  1,  1,  1,  1,  1,  1,  1].}
    \end{tabular}%
  \label{tab:timing}%
\end{table}%

\begin{table}[t]
  \centering
   \tabcolsep 7pt
  \caption{Compilers Elapsed Time Result}
    \begin{tabular}{|c|c|c|c|}
    \hline
    \multirow{2}[0]{*}{Compiler Name} &
      \multicolumn{3}{c|}{Circuit Complexity Level}
      \\ \cline{2-4}
     &
      Simple &
      Middle &
      Complex
      \\ \hline
    QUEKO \cite{tan2020optimal} &
      484.755s &
      5332.903s &
      Over 10 hours
      \\ \hline
    HA \cite{niu2020hardware} &
      4.765s &
      4.696s &
      4.201s
      \\ \hline
    OURS &
      0.008s &
      0.015s &
      0.020s\\
      
    Imp. (vs QUEKO) &
      60594.38$\times$ &
      355526.87$\times$ &
      >1800000$\times$ \\
      
    Imp. (vs HA) &
      595.63$\times$ &
      313.07$\times$ &
      210.05$\times$
      \\ \hline
    \end{tabular}%
  \label{tab:costtime}%
\end{table}%
\vspace{5pt}
\textit{A. Accuracy Comparison}
\vspace{5pt}

% Our algorithm is implemented in Python and the Qiskit
% version is 0.29.0. To evaluate our algorithms, we used Google Colab runs the experiments. 
We evaluate the performance of \net on both IBM Qiskit based simulator and the IBM Quantum Processor (i.e., the ibmq\_montreal backend with 27 qubits).

For the simulation, we set up different error rates with both bit-flip and phase-shift for the evaluation.
Table \ref{tab:timing} reports the results. 
% We evaluate the performance of the proposed error-aware implementation on quantum computation with with different error rates. 
% We call the best training result from the training set search as 'Searched\_Weight' and the initial weight as 'Baseline\_Weight'. 
% The obtained experimental results are shown in the Table \ref{tab:timing}, 
When the error rate is 0.0001, the obtained \net is exactly the same as the baseline model, because the baseline model is the one with the highest accuracy for perfect qubits, and when the error rate is small, it still performs better than other models.
However, when the error rate increased to larger than 0.0005, we observe that \net becomes different from the baseline model.
Specifically, \net achieves 1\%, 2\%, and 5\% accuracy improvements for the error rates set to 0.0005, 0.001, 0.01, respectively.
Furthermore, when the error increased to 0.05 and 0.1, the baseline model has less than 50\% accuracy, but \net identified by the proposed error-aware training framework can achieve 75\% accuracy, with 24\% and 28\% accuracy gain. 
From the above results, it is clear that the proposed training framework can work for different error settings.
In addition, with the increase of qubits' error, the model that works best for the perfect setting can easily become useless.
This emphasizes the importance of conducting error-aware learning.

% , the accuracy of 'Searched\_Weight' is improved by 1\% compared with 'Baseline\_Weight'. When the error rate is 0.001, the accuracy of 'Searched\_Weight' is increased by 2\% compared to 'Baseline\_Weight'. When the error rate rises to 0.01, the algorithm improves the accuracy rate by five percent. When the error rate is 0.05 and 0.1 respectively, the improvement of the accuracy of the searched 'Searched\_Weight' compared to 'Baseline\_Weight' reaches 24\% and 28\% respectively. There is an observation here that the larger the error rate is, the more significant the improvement of the algorithm on the accuracy rate is.
\vspace{5pt}
\textit{B. Efficiency Comparison}
\vspace{5pt}

We apply two existing compilers to be the baseline for efficiency comparison, including (1) Quantum Mapping Examples with Known Optimal (QUEKO) \cite{tan2020optimal} and (2) heuristic-based Hardware-Aware mapping algorithm (HA) \cite{niu2020hardware}.
Table \ref{tab:costtime} reports the comparison results on elapsed time.
Specifically, we run 
% , among them the heuristic-based Hardware-Aware mapping algorithm (HA) \cite{niu2020hardware} and Quantum Mapping Examples with Known Optimal (QUEKO) \cite{tan2020optimal}, which constructs benchmarks based on the inverse of the quantum device structure. The elapsed time results can be checked in the Table\ref{tab:costtime}. we run 
3 different circuits with circuit complex levels: simple, middle, and complex. 
QUEKO aims to explore the best solution for the general-purpose quantum circuit, and it takes more time than the other two compilers.
Specifically, QUEKO takes 484.775 seconds and 5332.903 seconds for simple and middle circuits, while it cannot optimize the complex circuit within 10 hours.
On the other hand, HA can complete all three circuits within 5 seconds.
Furthermore, the proposed application-specific compiler can complete the mapping within 0.1 seconds, achieving over 200$\times$ speedup over HA.
Results demonstrate that the proposed application-specific compiler can be efficiently integrated into the training framework.

% costs 484.755 as the simple circuit, 5332.904s in middle circuit, and over 36000s in complex circuit still not completed compiled. Since our quantum circuit generated by Baseline\_Weight is the complex circuit, we do not put QUEKO in next comparing experiment. HA remains a short elapsed time in all three circuits as 4.765s, 4.696s, and 4.201s. OURS show pretty short elapsed time as 0.008s, 0.015s, and 0.020s,respectively. Compare to the HA, we make improvement as 595.63 times reducing the elapsed time in simple model. Compare to the QUEKO, we make improvement as 355526.87 times reducing the elapsed time in middle model and over 1800000 times speeding up in complex model.

\vspace{5pt}
\textit{C. Accuracy, Efficiency, and Circuit Depth}
\vspace{5pt}

Finally, we execute benchmarks on ibmq\_montreal to have an end-to-end comparison between the solution obtained by using HA and \net obtained by the proposed approaches.
Table \ref{tab:eff} reports the results.
By using the baseline model and our compiling approach, the resultant system provides the accuracy of 74\%.
By applying \net, we can achieve improvement on accuracy to 75\% using HA and 80\% using our application-specific compiler.
One reason that we can obtain higher accuracy is because our approach can better optimize the circuit with less number of extra SWAP gates, where HA uses 43 SWAP gates while the figure is 12 in our approach.
In addition, we can see that our proposed compiler can achieve better efficiency over HA.

% the three compilers, the accuracy and extra gate number result shows in the Table \ref{tab:eff}. In this experiment, we run the quantum circuit generated by Baseline\_Weight with OURS on ibmq\_montreal. And learned the Searched\_Weight then using HA and OURS run the circuit generated by Searched\_Weight on ibmq\_montreal. Compared to HA, we achieve a better accuracy: ours is 80\% and HA is 75\%. We also achieved a shorter circuit length: we used 15 SWAP gates, HA used 43 SWAP gates. 

\begin{table}
    \centering
    \tabcolsep 0.8pt
    \caption{Compilers Results on ibmq\_montreal}
    \begin{tabular}{|c|c|c|c|c|}
    \hline
        Compiler Name & Model & Accuracy & Extra SWAP gate & Elapsed Time
        \\ \hline
        OURS & baseline & 74\% & 15 & 0.020s\\ \hline
        HA & \net & 75\% & 43 & 3.955s\\ \hline
        OURS & \net & 80\% & 12 & 0.014s\\ \hline
    
% \multicolumn{5}{l}{\footnotesize *HA and ours both on ibmq\_montreal machine.}
    \end{tabular}
\label{tab:eff}
\end{table}
% \begin{table}[htbp]
%   \centering
%   \caption{Compilers Result on ibmq\_montreal}
%     \begin{tabularx}{\linewidth}{lll}
%     \toprule
%     \thead{Compiler Name} &
%      \thead{Accuracy} &
%       \thead{Extra SWAP gate} &
%       \\
%     \midrule
%       HA &
%       75\% &
%       43 
%       \\
%     \midrule
%       OURS &
%       80\%&
%       15 
%       \\
%     \bottomrule
%     \end{tabularx}%
%   \label{tab:eff}%
% \end{table}%
\section{Conclusion}
In the NISQ era, noise is one critical issue needing to be resolved in quantum computing in order to make quantum computing practical.
It is still a long way to go before we can achieve perfect qubits, either achieving a breakthrough in material or realize quantum error correction.
In such a background, we proposed to design an application-specific optimization procedure to mitigate the effects from the qubits' noise.
Targeting Quantum Neural Network, we proposed the optimization framework to learn the error in the training procedure.
We further proposed an application-specific compiler to match the needs of high efficient mapping to support the lengthy training procedure.
Results on both simulation and actual quantum computers demonstrate that such an error-aware optimization can improve performance.

% modern quantum computers are not fully fault tolerant. A good quantum algorithm that can resist noise becomes just necessary. We propose an error-aware compiler with QNN algorithm that combines quantum mapping with quantum neural network exploration. The algorithm uses a quantum mapping approach to fix quantum circuits, followed by a quantum neural network to search for the best training weights to improve the accuracy. The experimental results show that our algorithm has good improvement on the accuracy rate, and the improvement is more significant as the error rate increases. We also have advantages in efficiency and accuracy compared with existing state-of-the-art compilers.

\section*{Acknowledgment}
Access to the IBM Q Network was obtained through the IBM Q Hub at NC State. Special thanks to Zhirui Hu for her extensive discussion and help with this work.

\printbibliography
\end{document}